\documentclass{IAU-TrA}
\usepackage{graphicx}

\title[ATOMIC AND MOLECULAR DATA] 
{}

\author[DIVISION~XII / COMMISSION~14 / WORKING GROUP] 
{ATOMIC DATA }

\pubyear{2012}
\volume{Volume XXVIIIA}
\pagerange{001--008}
\jname{Transactions IAU, Volume XXVIIIA}
\editors{Ian Corbett, ed.}
\begin{document}

\maketitle

{\bf

\large
\noindent
DIVISION XII / COMMISSION~14 / WORKING GROUP                \\ 
ATOMIC DATA                                                      \\

\normalsize

\begin{tabbing}
\hspace*{45mm}  \=                                                   \kill
CHAIR           \> Gillian Nave                                      \\
VICE-CHAIR      \> Glenn M. Wahlgren, Jeffrey R. Fuhr                \\
\end{tabbing}

\vspace{3mm}

\noindent
TRIENNIAL REPORT 2009-2012
}


This report summarizes laboratory measurements of atomic wavelengths,
energy levels, hyperfine and isotope structure, energy level
lifetimes, and oscillator strengths. Theoretical calculations of
lifetimes and oscillator strengths are also included. The bibliography
is limited to species of astrophysical interest. Compilations of
atomic data and internet databases are also included. Papers are
listed in the bibliography in alphabetical order, with a reference number in the
text.

\section{Energy levels, wavelengths, line classifications, and line structure}

Major analyses of wavelengths, energy levels and line classifications
have been published for {\bf V~I}  [\cite{15916EL}], {\bf Fe~II} [\cite{15000EL}],
[\cite{16577EL}], {\bf Cr~I} [\cite{14866EL}], {\bf Te~II} [\cite{15043EL}],
and {\bf Ho~II} [\cite{14526EL}] in the past three years. Wavelengths and energy levels
have also been measured in the symbiotic nova RR Telescopii for the
following species: Al VI, Ar III-V, C II, Ca V-VII, Cl IV, K IV-VI, Mg
V-VI, N II-III, Na~IV-VI, Ne III-V, O~IV, P~IV, S~IV-V, and Si~II-IV. 
Line identifications based on solar flares and active regions were
published for Ar XI,XIV, Ca XIV-XV, Fe XII, Fe XIII, Fe XVII, and Ni
XV [\cite{14040EL}].

Additional publications of
wavelengths, energy levels and line classifications include:

{\bf Al IV-XI }[\cite{17213EL}],
{\bf Ar II }[\cite{16364EL}],
{\bf Ba I }[\cite{15462EL}],
{\bf Ca I }[\cite{14943EL,17249EL}] IS [\cite{Salumbides}],
{\bf Ca XI }[\cite{14452EL}],
{\bf Ce II }[\cite{14909EL}],
{\bf Cl I }[\cite{14690EL}],
{\bf Co II} HFS [\cite{15052EL}],
{\bf Cr I} IS [\cite{14866EL}],
{\bf Cr III }[\cite{14097EL}],
{\bf D I }[\cite{15273EL}],
{\bf Dy I} HFS, IS [\cite{15609EL}],
{\bf Er I }[\cite{14408EL}], HFS, IS [\cite{14471EL}],
{\bf Er II }[\cite{14601EL}],
{\bf Eu I }[\cite{15917EL}],
{\bf Fe I} HFS, IS [\cite{15027EL}],
{\bf Fe II }[\cite{17184EL,16577EL,15000EL,14805EL}],
{\bf Fe VI-XIV }[\cite{14824EL}],
{\bf Fe VII-IX }[\cite{15032EL}],
{\bf Fe VII }[\cite{15001EL}],
{\bf Fe VIII }[\cite{15002EL,15110EL}],
{\bf Fe IX-XVI }[\cite{15556EL}],
{\bf Fe IX }[\cite{14442EL,15033EL}],
{\bf Fe X }[\cite{14158EL}],
{\bf Fe XI }[\cite{15194EL,15195EL,15216EL}],
{\bf Fe XII-XXII }[\cite{16371EL}],
{\bf Fe XIII }[\cite{17183EL}],
{\bf Fe XVII }[\cite{15022EL,16343EL}],
{\bf Gd I} IS [\cite{15199EL,14408EL}],
{\bf Ge I-II }[\cite{16364EL}],
{\bf H I} IS [\cite{15273EL,15495EL}],
{\bf Hg I} IS [\cite{15342EL}],
{\bf Ho II} HFS [\cite{14526EL}],
{\bf K I} IS, HFS [\cite{16262EL,17167EL}],
{\bf La I}  HFS [\cite{14451EL}],
{\bf La II} HFS [\cite{14268EL,14076EL,15025EL}],
{\bf Li I} IS [\cite{17230EL}],
{\bf Mg I} IS [\cite{14939EL,15487EL}],
{\bf Mg II} IS [\cite{14947EL}],
{\bf Mg III }[\cite{14994EL}],
{\bf Mg VII }[\cite{15556EL}],
{\bf Na I-XI }[\cite{13237EL}],
{\bf Na I  }[\cite{15340EL}],
{\bf Nb I} HFS [\cite{15295EL}],
{\bf Nb II} HFS [\cite{14265EL}],
{\bf O I }[\cite{14098EL}],
{\bf Os I} HFS [\cite{15076EL}],
{\bf Pb I-II} IS [\cite{14846EL,15501EL}],
{\bf Pr I-II} HFS [\cite{16267EL}],
{\bf S IX-XIII }[\cite{14522EL}],
{\bf S V }[\cite{14969EL}],
{\bf Si I} IS, HFS [\cite{16803EL}],
{\bf Si VII-IX }[\cite{15556EL}],
{\bf Sm I }[\cite{14154EL}],
{\bf Sn I }[\cite{12817EL}],
{\bf Ta I} HFS [\cite{14905EL}],
{\bf Ta II} HFS [\cite{14931EL}],
{\bf Tc I,II }[\cite{15237EL}],
{\bf Te II }[\cite{15043EL}],
{\bf Ti I} IS, HFS [\cite{14953EL}],
{\bf Ti II} IS, HFS [\cite{15202EL}],
{\bf Ti XIII }[\cite{14452EL}],
{\bf Tm I} HFS [\cite{14465EL}],
{\bf V I }[\cite{15916EL}],
{\bf V XXIII }[\cite{15082EL}],
{\bf Xe II }[\cite{17710EL}],
{\bf Zn II }[\cite{14800EL}]. \\
The references for elements heavier than Ni (Z$>$28) are limited to
the first three or four spectra only, these data being of most interest for
astronomical  spectroscopy.

Current analyses of neutral through doubly-ionized spectra are 
underway for iron-group spectra at the National Institute of Standards
and Technology (NIST) and the University of Wisconsin, USA and Imperial College,
London, UK. Work on rare-earth elements is being performed at the University of Wisconsin, USA;
Laboratoire Aim\'{ e} Cotton, Orsay, France; Observatoire de
Paris-Meudon, France; and the Institute of Spectroscopy, Troitsk,
Russia. Studies of more highly-ionized elements are being done using
electron beam ion traps at NIST, USA,  Lawrence Livermore National
Laboratory, USA, and Heidelberg, Germany, and with an accelerator in
Beijing, China.

\section{Wavelength standards}

Much of the work on wavelength standards during the period of this
report has focused on standards required for calibration of
astronomical spectrographs and for detecting possible changes in the
fine-structure constant during the history of the Universe.
Wavelengths emitted by a uranium-neon hollow cathode lamp suitable for
astronomical spectrograph calibration have been measured using Fourier
transform spectroscopy (FTS)
[\cite{17707EL}]. Updated wavelength standards suitable for detecting
changes in the fine-structure constant have been measured using
FTS for {\bf Mg I, Mg II, Ti~II, Cr~II, Mn~II, Fe
II} and {\bf Zn~II} [\cite{14800EL,16364EL}]. The most accurate
wavelength standards are now made using laser spectroscopy with a
laser frequency comb for calibration.  Frequency standards with
uncertainties of below 1 MHz have been published using this method for
{\bf Li I} [\cite{17230EL}], {\bf Mg II} [\cite{14947EL,15655EL}],
{\bf Ca I} [\cite{Salumbides}], and {\bf Ca II} 
[\cite{14130EL}]. Additional laser spectroscopy measurements include
the 546~nm line of {\bf $^{198}$Hg} [\cite{15342EL}], which was a
widely used wavelength standard for earlier studies. These
laser spectroscopy measurements have been used to validate the scale
of FTS measurements [\cite{16364EL}], putting all of these
measurements on the same wavelength scale. Ritz
wavelengths based on re-optimized energy levels of $^{198}$Hg are
found in [\cite{16369EL}].

\section{Transition probabilities}

The transition-probability data in the references in section \ref{TP_refs}
were obtained by both theoretical and experimental methods. The
references for elements heavier than Ni (Z$>$28) are limited to the
first three or four spectra only.

\section{Compilations, Reviews, Conferences}

Major compilations of wavelengths, energy levels or transition
probabilities have been published for the following elements: 
{\bf Al} [\cite{8199TP}],
{\bf Ar} [\cite{15349EL}],
{\bf B} [\cite{8702TP,12547EL,12548EL}],
{\bf Be} [\cite{8702TP}],
{\bf Cs} [\cite{15024EL}],
{\bf H I; D I; T I} [\cite{15291EL,8610TP}],
{\bf He} [\cite{8610TP}],
{\bf K} [\cite{8198TP}],
{\bf Li} [\cite{8610TP}],
{\bf Na}   [\cite{13237EL}],
{\bf S} [\cite{8595TP}],
{\bf Si} [\cite{8363TP}], and
{\bf Sr~I} [\cite{15350EL}]. Additional data can be found in {\em NIST
Atomic Transition Probabilities,} section of the Handbook of Chemistry
and Physics [\cite{CRC}].

Papers on atomic spectroscopic data are included in the proceedings of
the 10th International Conference on Atomic Spectra and Oscillator
Strengths [\cite{ASOS}], the 7th International Conference on Atomic and
Molecular data and their Applications [\cite{ICAMDATA}], and the 2010 NASA
Laboratory Astrophysics workshop [\cite{NASA_LAW}]. Additional
conferences including papers on atomic data include Atomic Processes
in Plasmas, the Congress of the European Group on Atomic Systems; and the
meeting of the Division of Atomic, Molecular and Optical Physics of
the American Physical Society.

\section{Databases}

The following databases of atomic spectra  at NIST 
have received significant updates since the last triennial report:

\begin{description}
\item{{\bf NIST Atomic Spectra Database:} \\}
{\tt http://www.nist.gov:/pml/data/asd.cfm } contains
critically compiled data on wavelengths, energy levels and
oscillators strengths.
\item{{\bf Ground Levels and Ionization Energies for the Neutral Atoms:} \\}
{\tt http://www.nist.gov/pml/data/ion\_energy.cfm}
\item{{\bf NIST Atomic Spectra Bibliographic Databases: } \\}
{\tt http://www.nist.gov/pml/data/asbib/index.cfm}\\
Consists of three databases of publications on atomic transition
probabilities, atomic energy levels and spectra, and atomic spectral
line broadening. 
\end{description}

Additional on-line databases including significant quantities of
atomic data include: 

\begin{description}
\sloppy
\item{{\bf The MCHF/MCDHF Collection on the Web}} (C.Froese Fischer
\textit{et al}.) at {\tt http://nlte.nist.gov/MCHF/index.html}
contains results of multi{}-configuration Hartree-Fock (MCHF)
or multi-configuration Dirac-Hartree-Fock (MCDHF) calculations for hydrogen and Li{}-like
through Ar{}-like ions, mainly for  Z $\leq$ 30.  Data for
fine{}-structure transitions are included. 

\item{{\bf The TOPbase and Opacity Projects}} include transition
probability and oscillator strength data for astrophysically abundant
ions (Z $\leq$ 26). A database is available at
{\tt http://cdsweb.u-strasbg.fr/topbase/topbase.html}

\item{{\bf  CHIANTI}}, an atomic database for spectroscopic diagnostics of
astrophysical plasmas at  {\tt http://www.chianti.rl.ac.uk/ }
contains atomic data and programs for computing spectra from
astrophysical plasmas, with the emphasis on highly-ionized atoms. 

\item{{\bf The Vienna Atomic Line Database (VALD)}}  web site
({\tt http://ams.astro.univie.ac.at/~vald/}) is a collection of atomic line
parameters of astronomical interest, with tools for selecting
subsets of astrophysical interest.

\item{{\bf The bibl database }} is a comprehensive bibliographic
database of experimental and theoretical papers on atomic spectroscopy,
with an emphasis on papers published since 1983.
It is available at {\tt http://das101.isan.troitsk.ru/bibl.htm}).

\end{description}

\section{Notes for References}
The references are identified by a
running number. This refers to the general reference list at the end of
this report, where the literature is ordered
alphabetically according to the first author. Each reference
contains one or more code letters indicating the method applied by the
authors, defined as follows:

\begin{description}
\vspace{0.1in}
\item{{\bf THEORETICAL METHODS:}}

\vspace{0.1in}
{\bf Q: } quantum mechanical calculations. \t {\bf QF: } Calculations of forbidden lines.
\vspace{0.1in}
\item{{\bf EXPERIMENTAL METHODS:}}

\vspace{0.1in}
\begin{tabular}{lll}
{\bf CL:} New classifications &
{\bf EL:}  Energy levels. &
{\bf WL:}  Wavelengths. \\
{\bf HFS:} Hyperfine structure. &
{\bf IS:}  Isotope structure. &
{\bf L:}   Lifetimes. \\
{\bf TE:} Experimental transition probabilities. \\
\end{tabular}
\vspace{0.1in}
\item{{\bf OTHER:}} 

\vspace{0.1in}
\begin{tabular}{lll}
{\bf CP: } Data compilations. &
{\bf R: }  Relative values only. &
{\bf F: } Forbidden lines. \\
\end{tabular}

\end{description}

\vspace{3mm}
{\hfill Gillian Nave }

{\hfill {\it chair of Working Group}}

\section{References on lifetimes and transition probabilities}\label{TP_refs}
\begin{minipage}[t]{1.8in}
Al II: \cite{8906TP} \\
Al III: \cite{8604TP} \\
Al IV: \cite{8843TP} \\
Al IX:   \cite{8850TP} \\
Al V-XII:   \cite{8841TP} \\
Al XI:  \cite{8594TP} \\

Ar V:   \cite{8656TP} \\
Ar VIII: \cite{8604TP} \\
Ar IX: \cite{8843TP} \\
Ar XVI: \cite{8594TP} \\

Au III:   \cite{8310TP} \\

B I:   \cite{8642TP} \\
B IV: \cite{8861TP} \\

Ba I:   \cite{15462EL} \\
Ba II:   \cite{8283TP,8893TP} \\

Be III: \cite{8861TP} \\

C I: \cite{8644TP} \\
C II:  \cite{8717TP,8159TP} \\
C V: \cite{8861TP} \\

Ca I:  \cite{14943EL} \\
Ca II:   \cite{8400TP,8893TP,8600TP} \\
Ca VIII:   \cite{8605TP} \\
Ca X: \cite{8604TP} \\
Ca XI: \cite{8843TP} \\
Ca XVII:  \cite{8524TP} \\
Ca XVIII: \cite{8594TP} \\

Cd II:  \cite{8893TP} \\

Ce I:   \cite{8607TP,8578TP} \\
Ce II: \cite{8369TP,14909EL} \\

Cl I:   \cite{14690EL,8279TP} \\
Cl VII: \cite{8604TP} \\
Cl VIII: \cite{8843TP} \\
Cl XV: \cite{8594TP} \\
\end{minipage}
\begin{minipage}[t]{2.2in}
Co XII: : \cite{8336TP} \\
Co XIII: \cite{8336TP} \\
Co XIV:  \cite{8336TP} \\
Co XV:  \cite{8362TP,8336TP} \\
Co XVII: \cite{8604TP} \\
Co XVIII: \cite{8843TP} \\
Co XXIII:    \cite{8850TP} \\

Cr II:  \cite{8580TP}, \cite{8716TP} \\
Cr VIII:   \cite{8667TP} \\
Cr XIV: \cite{8604TP} \\
Cr XV: \cite{8843TP} \\
Cr XXII:   \cite{8359TP} \\

Cu II:   \cite{8646TP,8275TP} \\

Er I:  \cite{8846TP,8807TP} \\
Er II: \cite{8512TP,14601EL} \\

Eu I:  \cite{17164EL} \\
Eu III: \cite{8157TP} \\

F VII: \cite{8756TP} \\
F VIII: \cite{8665TP} \\
F IX: \cite{8825TP} \\

Fe II: \cite{16577EL,15000EL,8732TP,8677TP,8561TP,8690TP} \\
Fe III: \cite{8274TP,8514TP,8525TP} \\
Fe IV: \cite{8225TP,8746TP,8341TP,8277TP} \\
Fe VI: \cite{8375TP} \\
Fe VII: \cite{8316TP,15001EL} \\
Fe VIII: \cite{15110EL,15002EL} \\
Fe IX: \cite{15033EL} \\
Fe X: \cite{8645TP} \\
Fe XI: \cite{8336TP,15195EL,15194EL} \\
Fe XII: \cite{8660TP,8921TP,8281TP,8336TP} \\
Fe XIII: \cite{8715TP,8336TP} \\
Fe XIV: \cite{8805TP,8356TP,8660TP,8921TP,8336TP} \\
Fe XV:  \cite{8929TP,8606TP} \\
Fe XVI: \cite{8604TP,8521TP} \\
Fe XVII:  \cite{8843TP,8788TP} \\
Fe XVII-XXV: \cite{8523TP} \\
Fe XIX: \cite{8321TP,8688TP,8925TP} \\
Fe XX:   \cite{8688TP} \\
Fe XXII: \cite{8850TP,8338TP,8691TP} \\
Fe XXIII: \cite{8331TP} \\
Fe XXIV: \cite{8916TP,8176TP} \\
Fe XXVI: \cite{8158TP,8727TP} \\
\end{minipage}
\begin{minipage}[t]{2in}
Ga I:   \cite{8573TP} \\

Gd I: L  \cite{8845TP,8877TP} \\

Ge IV:    \cite{8910TP} \\

He I:     \cite{8332TP,8875TP} \\

Hf I,III:  \cite{8615TP} \\

Hg II: \cite{8893TP} \\

In I:     \cite{8913TP} \\

K I:   \cite{8230TP,8173TP,8349TP} \\
K II:   \cite{8714TP} \\
K IX: \cite{8604TP} \\
K X: \cite{8843TP} \\
K XVII: \cite{8594TP} \\

Kr II:   \cite{8776TP} \\

La I:   \cite{8858TP,8233TP,8152TP} \\
La II:   \cite{8648TP} \\

Li I-II:   \cite{8643TP} \\
Li II: \cite{8861TP} \\

Mg I:   \cite{8906TP,8930TP,8577TP} \\
Mg II:   \cite{15097EL,8604TP} \\
Mg III: \cite{8843TP} \\
Mg IX:  \cite{8405TP,8319TP} \\
Mg X:  \cite{8594TP} \\

Mn I: \cite{8842TP} \\
Mn XV: \cite{8604TP} \\
Mn XVI: \cite{8843TP} \\
Mn XXI: \cite{8850TP} \\

Mo II: \cite{8750TP} \\

N I-VII: \cite{8678TP} \\
N I: \cite{8186TP,8644TP,8783TP} \\
N II: \cite{8644TP,8753TP,8851TP} \\
N III: \cite{8717TP} \\
N V: \cite{8756TP} \\
N VI: \cite{8665TP} \\
N VII: \cite{8825TP} \\
\end{minipage}
\newpage
\begin{minipage}[t]{2in}
Na I: \cite{8153TP} \\
Na II:  \cite{8843TP} \\
Na IX: \cite{8756TP,8594TP} \\
Na X: \cite{8665TP} \\
Na XI: \cite{8825TP} \\

Nb I:  \cite{8874TP} \\
Nb II: \cite{8731TP,14265EL} \\
Nb III:   \cite{8731TP} \\

Nd II:   \cite{8673TP} \\

Ne I:   \cite{8378TP} \\
Ne II:   \cite{8193TP,8917TP} \\
Ne VI:   \cite{8329TP} \\
Ne VIII:  \cite{8756TP} \\
Ne X:   \cite{8825TP} \\

Ni XI:    \cite{8909TP} \\
Ni XIII:  \cite{8336TP} \\
Ni XIV:  \cite{8689TP,8336TP} \\
Ni XV:  \cite{8336TP} \\
Ni XVI:  \cite{8336TP} \\
Ni XVII:    \cite{8872TP} \\
Ni XIX \cite{8843TP} \\
Ni XIX-XXVII:   \cite{8398TP} \\
Ni XXIII:   \cite{8932TP} \\
Ni XXIV:   \cite{8850TP} \\
Ni XXV:    \cite{8320TP,8630TP} \\

O I:   \cite{8234TP,8254TP,8761TP,8518TP} \\
O II:   \cite{8613TP,8644TP,8625TP,8785TP} \\
O III:   \cite{8639TP} \\
O IV:    \cite{8317TP} \\
O IV:   \cite{8717TP} \\
O VII:    \cite{8322TP} \\
O VIII:   \cite{8825TP} \\

P II:    \cite{8890TP,8275TP} \\
P IV: \cite{8906TP} \\
P V: \cite{8604TP} \\
P VI: \cite{8843TP} \\
P XII:   \cite{8853TP} \\
P XIII: \cite{8594TP} \\

Pb III:   \cite{8575TP} \\

Pr II-III:   \cite{8528TP} \\

Pt II:  \cite{8261TP} \\
\end{minipage}
\begin{minipage}[t]{2in}
Rb I:    \cite{8886TP} \\

Rh II:   \cite{8920TP} \\

Ru I:   \cite{8633TP} \\
Ru II-III:   \cite{8620TP} \\

S I:   \cite{8867TP,8901TP,8169TP} \\
S II:    \cite{8758TP} \\
S V: \cite{8906TP} \\
S VI: \cite{8604TP} \\
S VII: \cite{8843TP} \\
S XI-XV:   \cite{8517TP} \\
S XIII:   \cite{8164TP,14359EL}\\
S XIV: \cite{8594TP,8619TP} \\
S XV:   \cite{8231TP} \\

Sb I:  \cite{8803TP} \\

Sc II:    \cite{8156TP,8309TP} \\
Sc II:  \cite{8580TP,8156TP,8309TP} \\
Sc III:    \cite{8371TP,8612TP} \\
Sc XI: \cite{8604TP} \\
Sc XII: \cite{8843TP} \\

Si I:   \cite{8247TP,8358TP} \\
Si II-IV:   \cite{8700TP} \\
Si II:   \cite{8686TP,8700TP} \\
Si III: \cite{8906TP} \\
Si IV: \cite{8604TP} \\
Si V: \cite{8843TP} \\
Si VI:   \cite{8147TP} \\
Si X:   \cite{8593TP} \\
Si XI:   \cite{8823TP} \\
Si XII: \cite{8594TP} \\
Si XII-XIV: \cite{8822TP} \\

Sm I:   \cite{8742TP,8795TP,8808TP} \\

Sn I:   \cite{8743TP,8647TP,8790TP,12817EL} \\
Sn II:   \cite{8718TP} \\
Sn III:   \cite{8778TP} \\

Sr II:  \cite{8893TP} \\

Ta II:    \cite{8406TP} \\
Ta III:   \cite{9991TP}\\

Tb I,II:   \cite{8804TP} \\
\end{minipage}
\begin{minipage}[t]{2in}
Ti II:    \cite{8335TP,8227TP} \\
Ti III-IV:   \cite{8333TP} \\
Ti IV:    \cite{8228TP,8649TP,8779TP,8370TP} \\
Ti X:   \cite{8762TP,8789TP} \\
Ti XII: \cite{8604TP} \\
Ti XIII: \cite{8843TP} \\

V XI:   \cite{8749TP} \\
V XIII: \cite{8604TP} \\
V XIV: \cite{8843TP} \\

W II: \cite{8361TP} \\
W III:  \cite{9992TP} \\

Xe II:   \cite{17710EL} \\

Y II-III:  \cite{8911TP} \\
Y III:   \cite{8371TP,8612TP} \\

Yb I:   \cite{8870TP} \\
Yb II:   \cite{8766TP} \\

Zn I:   \cite{8623TP,8235TP} \\
Zn II:  \cite{8367TP,8623TP} \\

Zr I:   \cite{8588TP,8871TP} \\
\end{minipage}

\begin{thebibliography}{}

\bibitem[1]{ICAMDATA}
[1] 7th International Conference on Atomic and Molecular Data and their Applications: 2010,
\newblock AIP Conference Proceedings Volume 1344 

\bibitem[2]{ASOS}
[2] 10th International Colloquium on Atomic Spectra and Oscillator
Strengths for Astrophysical and Laboratory Plasmas: 2011
\newblock Canadian J. Physics, 89

\bibitem[3]{NASA_LAW}
[3] The 2010 NASA Laboratory Astrophysics Workshop: 2011, ed. D.R. Schultz, Oak Ridge National Laboratory 
\newblock http://www-cfadc.phy.ornl.gov/nasa\_law/proceedings.html

\bibitem[4]{8158TP}
[4] Aggarwal, K.~M., Hamada, K., Igarashi, A., Jonauskas, V., Keenan, F.~P., and
  Nakazaki, S.: 2008,
\newblock Astron. Astrophys. 484, 879,
\newblock {\bf  Q, QF }

\bibitem[5]{8667TP}
[5] Aggarwal, K.~M., Kato, T., Keenan, F.~P., and Murakami, I.: 2009a,
\newblock Astron. Astrophys. 506, 1501,
\newblock {\bf  Q, QF }

\bibitem[6]{8861TP}
[6] Aggarwal, K.~M., Kato, T., Keenan, F.~P., and Murakami, I.: 2011,
\newblock Phys. Scr. 83, 015302,
\newblock {\bf  Q, QF }

\bibitem[7]{8322TP}
[7] Aggarwal, K.~M. and Keenan, F.~P.: 2008a,
\newblock Astron. Astrophys. 489, 1377,
\newblock {\bf  Q, QF }

\bibitem[8]{8317TP}
[8] Aggarwal, K.~M. and Keenan, F.~P.: 2008b,
\newblock Astron. Astrophys. 486, 1053,
\newblock {\bf  Q, QF }

\bibitem[9]{8822TP}
[9] Aggarwal, K.~M. and Keenan, F.~P.: 2010,
\newblock Phys. Scr. 82, 065302,
\newblock {\bf  Q, QF }

\bibitem[10]{8665TP}
[10] Aggarwal, K.~M., Keenan, F.~P., and Heeter, R.~F.: 2009b,
\newblock Phys. Scr. 80, 045301,
\newblock {\bf  Q, QF }

\bibitem[11]{8825TP}
[11] Aggarwal, K.~M., Keenan, F.~P., and Heeter, R.~F.: 2010a,
\newblock Phys. Scr. 82, 015006,
\newblock {\bf  Q, QF }

\bibitem[12]{8756TP}
[12] Aggarwal, K.~M., Keenan, F.~P., and Heeter, R.~F.: 2010b,
\newblock Phys. Scr. 81, 015303,
\newblock {\bf  Q, QF  }

\bibitem[13]{14465EL}
[13] Akimov, A.~V., Chebakov, K.~Y., Tolstikhina, I.~Y., Sokolov, A.~V., Rodionov,
  P.~B., Kanorsky, S.~I., Sorokin, V.~N., and Kolachevsky, N.~N.: 2008,
\newblock Quantum Electron. 38, 961,
\newblock {\bf  HFS }

\bibitem[14]{14800EL}
[14] Aldenius, M.: 2009,
\newblock Phys. Scr. T134, 014008,
\newblock {\bf  WL, CL  }

\bibitem[15]{14943EL}
[15] Aldenius, M., Lundberg, H., and Blackwell-Whitehead, R.: 2009a,
\newblock Astron. Astrophys. 502, 989,
\newblock {\bf  WL, CL, TE, EL  }

\bibitem[16]{8332TP}
[16] Alexander, S.~A. and Coldwell, R.~L.: 2008,
\newblock Int. J. Quantum Chem. 108, 2813,
\newblock {\bf   Q, QF }

\bibitem[17]{8743TP}
[17] Alonso-Medina, A.: 2010,
\newblock Spectrochim. Acta, Part B 65, 158,
\newblock {\bf  TE }

\bibitem[18]{8575TP}
[18] Alonso-Medina, A., Col\'on, C., and Zan\'on, A.: 2009,
\newblock Mon. Not. R. Astron. Soc. 395, 567,
\newblock {\bf  Q }

\bibitem[19]{15199EL}
[19] Ankush, B.~K. and Deo, M.~N.: 2010,
\newblock Phys. Scr. 81, 055301,
\newblock {\bf  IS, EL }

\bibitem[20]{8230TP}
[20] J.M.P. Serrao, J. M. P.~S.: 2008,
\newblock J. Quant. Spectrosc. Radiat. Transfer 109, 453,
\newblock {\bf  Q }

\bibitem[21]{15495EL}
[21] Arnoult, O., Nez, F., Julien, L., and Biraben, F.: 2010,
\newblock Eur. Phys. J. D 60, 243,
\newblock {\bf  WL, EL }

\bibitem[22]{8643TP}
[22] Ate{\c s}, {\c S}. and {\c C}elik, G.: 2009,
\newblock Acta Phys. Pol. A 116(2), 169,
\newblock {\bf  Q }

\bibitem[23]{8613TP}
[23] Ate{\c s}, {\c S}., Tekeli, G., {\c C}elik, G., Akin, E., and Ta{\c s}er, M.:
  2009,
\newblock Eur. Phys. J. D 54, 21,
\newblock {\bf  Q }

\bibitem[24]{8234TP}
[24] Bac{\l}awski, A.: 2008a,
\newblock J. Quant. Spectrosc. Radiat. Transfer 109, 1986,
\newblock {\bf  TE, R }

\bibitem[25]{8378TP}
[25] Bac{\l}awski, A.: 2008b,
\newblock J. Phys. B 41, 225701,
\newblock {\bf  TE }

\bibitem[26]{8867TP}
[26] Bac{\l}awski, A.: 2011,
\newblock Eur. Phys. J. D 61, 327,
\newblock {\bf  TE, R }

\bibitem[27]{8186TP}
[27] Bac{\l}awski, A. and Musielok, J.: 2008a,
\newblock J. Quant. Spectrosc. Radiat. Transfer 109, 2537,
\newblock {\bf  TE, R }

\bibitem[28]{8254TP}
[28] Bac{\l}awski, A. and Musielok, J.: 2008b,
\newblock Spectrochim. Acta, Part B 63, 1315,
\newblock {\bf  TE }

\bibitem[29]{8644TP}
[29] Bac{\l}awski, A. and Musielok, J.: 2009,
\newblock Acta Phys. Pol. A 116(2), 176,
\newblock {\bf  CP }

\bibitem[30]{8901TP}
[30] Bac{\l}awski, A. and Musielok, J.: 2011,
\newblock J. Phys. B 44, 135002,
\newblock {\bf  TE, R }

\bibitem[31]{8375TP}
[31] Ballance, C.~P. and Griffin, D.~C.: 2008,
\newblock J. Phys. B 41, 195205,
\newblock {\bf  Q }

\bibitem[32]{14451EL}
[32] Ba{\c s}ar, G., Ba{\c s}ar, G., and Kr\"oger, S.: 2009,
\newblock Opt. Commun. 282, 562,
\newblock {\bf  HFS, EL }

\bibitem[33]{15076EL}
[33] Ba{\c s}ar, G., Ba{\c s}ar, G., Kr\"oger, S., and Guth\"ohrlein, G.~H.: 2010,
\newblock J. Phys. B 43, 074008,
\newblock {\bf  HFS }

\bibitem[34]{14947EL}
[34] Batteiger, V., Kn\"unz, S., Herrmann, M., Saathoff, G., Sch\"ussler, H.~A.,
  Bernhardt, B., Wilken, T., Holzwarth, R., H\"ansch, T.~W., and Udem, T.:
  2009,
\newblock Phys. Rev. A 80, 022503,
\newblock {\bf  IS, WL }


\bibitem[35]{8580TP}
[35] Bautista, M.~A., Ballance, C., Gull, T.~R., Hartman, H., Lodders, K.,
  Mart\'inez, M., and Mel\'endez, M.: 2009a,
\newblock Mon. Not. R. Astron. Soc. 393, 1503,
\newblock {\bf  QF }

\bibitem[36]{8686TP}
[36] Bautista, M.~A., Quinet, P., Palmeri, P., Badnell, N.~R., Dunn, J., and Arav,
  N.: 2009b,
\newblock Astron. Astrophys. 508, 1527,
\newblock {\bf  Q }

\bibitem[37]{17167EL}
[37] Behrle, A., Koschorreck, M., and K\"ohl, M.: 2011,
\newblock Phys. Rev. A 83, 052507,
\newblock {\bf  IS, HFS }

\bibitem[38]{15052EL}
[38] Bergemann, M., Pickering, J.~C., and Gehren, T.: 2010,
\newblock Mon. Not. R. Astron. Soc. 401, 1334,
\newblock {\bf   HFS }

\bibitem[39]{8909TP}
[39] Bhatia, A.~K. and Landi, E.: 2011a,
\newblock At. Data Nucl. Data Tables 97, 50,
\newblock {\bf  Q, QF }

\bibitem[40]{8872TP}
[40] Bhatia, A.~K. and Landi, E.: 2011b,
\newblock At. Data Nucl. Data Tables 97, 189,
\newblock {\bf  Q, QF }

\bibitem[41]{8911TP}
[41] Bi\'emont, E., Blagoev, K., Engstr\"om, L., Hartman, H., Lundberg, H.,
  Malcheva, G., Nilsson, H., Whitehead, R.~B., Palmeri, P., and Quinet, P.:
  2011,
\newblock Mon. Not. R. Astron. Soc. 414, 3350,
\newblock {\bf  L, Q }

\bibitem[42]{8842TP}
[42] Blackwell-Whitehead, R., Pavlenko, Y.~V., Nave, G., Pickering, J.~C., Jones, H.
  R.~A., Lyubchik, Y., and Nilsson, H.: 2011,
\newblock Astron. Astrophys. 525, p. A44,
\newblock {\bf  TE }

\bibitem[43]{8645TP}
[43] Brenner, G., L\'opez-Urrutia, J. R.~C., Bernitt, S., Fischer, D., Ginzel, R.,
  Kubi{\v c}ek, K., M\"ackel, V., Mokler, P.~H., Simon, M.~C., and Ullrich, J.:
  2009,
\newblock Astrophys. J. 703, 68,
\newblock {\bf  LF }

\bibitem[44]{14690EL}
[44] Bridges, J.~M. and Wiese, W.~L.: 2008a,
\newblock Phys. Rev. A 78, 062508,
\newblock {\bf  TE }

\bibitem[45]{8783TP}
[45] Bridges, J.~M. and Wiese, W.~L.: 2010,
\newblock Phys. Rev. A 82, 024502,
\newblock {\bf  TE }

\bibitem[46]{14994EL}
[46] Brown, C.~M., Kramida, A.~E., Feldman, U., and Reader, J.: 2009a,
\newblock Phys. Scr. 80, 065302,
\newblock {\bf  EL, CL, WL }

\bibitem[47]{8890TP}
[47] Brown, J.~B., Brown, M.~S., Cheng, S., Curtis, L.~J., Ellis, D.~G., Federman,
  S.~R., and Irving, R.~E.: 2011,
\newblock Can. J. Phys. 89, 413,
\newblock {\bf  Q, R }

\bibitem[48]{8646TP}
[48] Brown, M.~S., Federman, S.~R., Irving, R.~E., Cheng, S., and Curtis, L.~J.:
  2009b,
\newblock Astrophys. J. 702, 880,
\newblock {\bf   L }

\bibitem[49]{8321TP}
[49] Butler, K. and Badnell, N.~R.: 2008,
\newblock Astron. Astrophys. 489, 1369,
\newblock {\bf  Q }

\bibitem[50]{17184EL}
[50] Castelli, F., Johansson, S., and Hubrig, S.: 2008,
\newblock J. Phys.: Conf. Ser. 130, 012003,
\newblock {\bf  CL, WL }

\bibitem[51]{16577EL}
[51] Castelli, F. and Kurucz, R.~L.: 2010a,
\newblock Astron. Astrophys. 520, p. A57,
\newblock {\bf   EL, CL, Q }

\bibitem[52]{15000EL}
[52] Castelli, F., Kurucz, R.~L., and Hubrig, S.: 2009a,
\newblock Astron. Astrophys. 508, 401,
\newblock {\bf  CL, WL, EL, Q }

\bibitem[53]{8153TP}
[53] {\c C}elik, G. and Ate{\c s}, {\c S}.: 2008a,
\newblock Acta Phys. Pol. A 113(6), 1619,
\newblock {\bf  Q }

\bibitem[54]{8173TP}
[54] {\c C}elik, G. and Ate{\c s}, {\c S}.: 2008b,
\newblock Can. J. Phys. 86, 487,
\newblock {\bf  Q }

\bibitem[55]{8727TP}
[55] Chen, C.-Y., Wang, K., Huang, M., Wang, Y.-S., and Zou, Y.-M.: 2010,
\newblock J. Quant. Spectrosc. Radiat. Transfer 111, 843,
\newblock {\bf  Q, QF }

\bibitem[56]{8906TP}
[56] Cheng, C., Gao, X., Qing, B., Zhang, X.-L., and Li, J.-M.: 2011,
\newblock Chin. Phys. B 20, 033103,
\newblock {\bf  Q }

\bibitem[57]{14464EL}
[57] Chwalla, M., Benhelm, J., Kim, K., Kirchmair, G., Monz, T., Riebe, M.,
  Schindler, P., Villar, A.~S., H\"ansel, W., Roos, C.~F., Blatt, R., Abgrall,
  M., Santarelli, G., Rovera, G.~D., and Laurent, P.: 2009,
\newblock Phys. Rev. Lett. 102, 023002,
\newblock {\bf  WL  }

\bibitem[58]{8778TP}
[58] Col\'on, C. and Alonso-Medina, A.: 2010,
\newblock J. Phys. B 43, 165001,
\newblock {\bf  Q }

\bibitem[59]{8607TP}
[59] Curry, J.~J.: 2009,
\newblock J. Phys. D 42, 135205,
\newblock {\bf  TE  }

\bibitem[60]{17249EL}
[60] Dammalapati, U., Norris, I., Burrows, C., and Riis, E.: 2011,
\newblock Phys. Rev. A 83, 062513,
\newblock {\bf  CL, WL }

\bibitem[61]{8225TP}
[61] Deb, N.~C. and Hibbert, A.: 2008a,
\newblock J. Phys. B 41, 081007,
\newblock {\bf  Q  }

\bibitem[62]{8169TP}
[62] Deb, N.~C. and Hibbert, A.: 2008b,
\newblock At. Data Nucl. Data Tables 94, 561,
\newblock {\bf  Q }

\bibitem[63]{8274TP}
[63] Deb, N.~C. and Hibbert, A.: 2008c,
\newblock J. Phys.: Conf. Ser. 130, 012006,
\newblock {\bf  Q, QF }

\bibitem[64]{8514TP}
[64] Deb, N.~C. and Hibbert, A.: 2009a,
\newblock J. Phys. B 42, 065003,
\newblock {\bf  QF }

\bibitem[65]{8525TP}
[65] Deb, N.~C. and Hibbert, A.: 2009b,
\newblock At. Data Nucl. Data Tables 95, 184,
\newblock {\bf  Q }

\bibitem[66]{8732TP}
[66] Deb, N.~C. and Hibbert, A.: 2010a,
\newblock Astrophys. J. 711, L104,
\newblock {\bf  QF }

\bibitem[67]{8746TP}
[67] Deb, N.~C. and Hibbert, A.: 2010b,
\newblock At. Data Nucl. Data Tables 96, 358,
\newblock {\bf  Q }

\bibitem[68]{8335TP}
[68] Deb, N.~C., Hibbert, A., Felfli, Z., and Msezane, A.~Z.: 2009,
\newblock J. Phys. B 42, 015701,
\newblock {\bf  QF }

\bibitem[69]{15097EL}
[69] Destree, J.~D., Williamson, K.~E., and Snow, T.~P.: 2010a,
\newblock Astrophys. J. 712, L48,
\newblock {\bf  CL, WL, TE }

\bibitem[70]{8367TP}
[70] Dixit, G., Nataraj, H.~S., Sahoo, B.~K., Chaudhuri, R.~K., and Majumder, S.:
  2008,
\newblock J. Phys. B 41, 025001,
\newblock {\bf  Q, QF }

\bibitem[71]{8623TP}
[71] Dixit, G., Sahoo, B.~K., Chaudhuri, R.~K., and Majumder, S.: 2009,
\newblock J. Phys. B 42, 165702,
\newblock {\bf  Q, QF }

\bibitem[72]{8700TP}
[72] Djeni{\v z}e, S., Sre\'ckovi\'c, A., and Bukvi\'c, S.: 2010,
\newblock Spectrochim. Acta, Part B 65, 61,
\newblock {\bf  TE, R }

\bibitem[73]{8320TP}
[73] Duan, B., Bari, M.~A., Zhong, J.~Y., Yan, J., Li, Y.~M., and Zhang, J.: 2008,
\newblock Astron. Astrophys. 488, 1155,
\newblock {\bf  Q, QF }

\bibitem[74]{8910TP}
[74] Dutta, N.~N. and Majumder, S.: 2011,
\newblock Astrophys. J. 737, 25,
\newblock {\bf  Q, QF }

\bibitem[75]{8517TP}
[75] Fan, Q., Liao, Z.~J., Yang, J.~H., and Zhang, J.~P.: 2009,
\newblock Phys. Scr. 79, 015301,
\newblock {\bf  Q }

\bibitem[76]{8275TP}
[76] Federman, S.~R., Curtis, L.~J., Brown, M., Cheng, S., Irving, R.~E., Torok, S.,
  and Schectman, R.~M.: 2008,
\newblock J. Phys.: Conf. Ser. 130, 012007,
\newblock {\bf  TE, R, L }

\bibitem[77]{8858TP}
[77] Feng, Y.-Y., Zhang, W., Kuang, B., Ning, L.-L., Jiang, Z.-K., and Dai, Z.-W.:
  2011,
\newblock J. Opt. Soc. Am. B 28, 543,
\newblock {\bf  L }

\bibitem[78]{8804TP}
[78] Feng, Y.-Y., Zhang, W., Ning, L.-L., Kuang, B., Sun, G.-J., and Dai, Z.-W.:
  2010,
\newblock J. Phys. B 43, 225001,
\newblock {\bf  L }

\bibitem[79]{8193TP}
[79] Fischer, C.~F. and Ralchenko, Y.: 2008,
\newblock Int. J. Mass Spectrom. 271, 85,
\newblock {\bf  Q }

\bibitem[80]{8341TP}
[80] Fischer, C.~F., Rubin, R.~H., and Rodr\'iguez, M.: 2008,
\newblock Mon. Not. R. Astron. Soc. 391, 1828,
\newblock {\bf  QF }

\bibitem[81]{8639TP}
[81] Fischer, C.~F., Tachiev, G., Rubin, R.~H., and Rodr\'iguez, M.: 2009,
\newblock Astrophys. J. 703, 500,
\newblock {\bf  Q, QF  }

\bibitem[82]{9991TP}
[82] Fivet, V., Bi\'emont, E., Engstr\"om, L., Lundberg, H., Nilsson, H., Palmeri, P., and Quinet, P.:
  2008,
\newblock J. Phys. B 41, 015702,
\newblock {\bf  L, Q }

\bibitem[83]{8633TP}
[83] Fivet, V., Quinet, P., Palmeri, P., Bi\'emont, E., Asplund, M., Grevesse, N.,
  Sauval, A.~J., Engstr\"om, L., Lundberg, H., Hartman, H., and Nilsson, H.:
  2009,
\newblock Mon. Not. R. Astron. Soc. 396, 2124,
\newblock {\bf  L, Q }

\bibitem[84]{8702TP}
[84] Fuhr, J.~R. and Wiese, W.~L.: 2010,
\newblock J. Phys. Chem. Ref. Data 39, 013101,
\newblock {\bf  CP }

\bibitem[85]{CRC}
[85] Fuhr, J.~R. and Wiese, W.~L.: 2011,
\newblock CRC Handbook of Chemistry and Physics, 92nd 
Edition, (Ed. D.~R.~Lide) CRC Press, Boca Raton, FL,

\bibitem[86]{14268EL}
[86] Furmann, B., Elantkowska, M., Stefa\'nska, D., Ruczkowski, J., and
  Dembczy\'nski, J.: 2008a,
\newblock J. Phys. B 41, 235002,
\newblock {\bf  HFS }

\bibitem[87]{14076EL}
[87] Furmann, B., Ruczkowski, J., Stefa\'nska, D., Elantkowska, M., and
  Dembczy\'nski, J.: 2008b,
\newblock J. Phys. B 41, 215004,
\newblock {\bf  HFS }

\bibitem[88]{15025EL}
[88] Furmann, B., Stefa\'nska, D., and Dembczy\'nski, J.: 2010,
\newblock J. Phys. B 43, 015001,
\newblock {\bf  EL,CL, WL, HFS }

\bibitem[89]{16267EL}
[89] Gamper, B., Uddin, Z., Jahangir, M., Allard, O., Kn\"ockel, H., Tiemann, E.,
  and Windholz, L.: 2011,
\newblock J. Phys. B 44, 045003,
\newblock {\bf  CL, WL, EL,HFS }

\bibitem[90]{8678TP}
[90] Garc\'ia, J., Kallman, T.~R., Witthoeft, M., Behar, E., Mendoza, C., Palmeri,
  P., Quinet, P., Bautista, M.~A., and Klapisch, M.: 2009,
\newblock Astrophys. J., Suppl. Ser. 185, 477,
\newblock {\bf  Q }

\bibitem[91]{8761TP}
[91] Gattinger, R.~L., Lloyd, N.~D., Bourassa, A.~E., Degenstein, D.~A., McDade,
  I.~C., and Llewellyn, E.~J.: 2009,
\newblock Can. J. Phys. 87, 1133,
\newblock {\bf   TE-R-F }

\bibitem[92]{8400TP}
[92] Gerritsma, R., Kirchmair, G., Z\"ahringer, F., Benhelm, J., Blatt, R., and
  Roos, C.~F.: 2008,
\newblock Eur. Phys. J. D 50, 13,
\newblock {\bf  TE }

\bibitem[93]{15082EL}
[93] Gillaspy, J.~D., Chantler, C.~T., Paterson, D., Hudson, L.~T., Serpa, F.~G.,
  and Tak\'acs, E.: 2010,
\newblock J. Phys. B 43, 074021,
\newblock {\bf  CL, EL,WL }

\bibitem[94]{16343EL}
[94] Gillaspy, J.~D., Lin, T., Tedesco, L., Tan, J.~N., Pomeroy, J.~M., Laming,
  J.~M., Brickhouse, N., Chen, G.-X., and Silver, E.: 2011,
\newblock Astrophys. J. 728, 132,
\newblock {\bf  WL }

\bibitem[95]{14905EL}
[95] G{\l}owacki, P., Uddin, Z., Guth\"ohrlein, G.~H., Windholz, L., and
  Dembczy\'nski, J.: 2009,
\newblock Phys. Scr. 80, 025301,
\newblock {\bf  CL, WL, EL, HFS }

\bibitem[96]{17213EL}
[96] Gu, M.~F., Beiersdorfer, P., and Lepson, J.~K.: 2011,
\newblock Astrophys. J. 732, 91,
\newblock {\bf  CL, WL }

\bibitem[97]{14154EL}
[97] Guan, F., Dai, C.-J., and Zhao, H.-Y.: 2008,
\newblock Chin. Phys. B 17, 3655,
\newblock {\bf  EL, CL, WL }

\bibitem[98]{8362TP}
[98] Gupta, G.~P. and Msezane, A.~Z.: 2008,
\newblock Eur. Phys. J. D 49, 157,
\newblock {\bf  Q }

\bibitem[99]{8749TP}
[99] Gupta, G.~P. and Msezane, A.~Z.: 2010,
\newblock Phys. Scr. 81, 045302,
\newblock {\bf  Q }

\bibitem[100]{8677TP}
[100] Gurell, J., Hartman, H., Blackwell-Whitehead, R., Nilsson, H., B\"ackstr\"om,
  E., Norlin, L.~O., Royen, P., and Mannervik, S.: 2009a,
\newblock Astron. Astrophys. 508, 525,
\newblock {\bf  L, F }

\bibitem[101]{8716TP}
[101] Gurell, J., Nilsson, H., Engstr\"om, L., Lundberg, H., Blackwell-Whitehead, R.,
  Nielsen, K.~E., and Mannervik, S.: 2010,
\newblock Astron. Astrophys. 511, p. A68,
\newblock {\bf  L, TE }

\bibitem[102]{14526EL}
[102] Gurell, J., Wahlgren, G.~M., Nave, G., and Wyart, J.-F.: 2009b,
\newblock Phys. Scr. 79, 035306,
\newblock {\bf  CL, WL, HFS, EL }

\bibitem[103]{8147TP}
[103] Hamdi, R., Nessib, N.~B., Milovanovi\'c, N., Popovi\'c, L.~{\v C}.,
  Dimitrijevi\'c, M.~S., and Sahal-Br\'echot, S.: 2008,
\newblock Mon. Not. R. Astron. Soc. 387, 871,
\newblock {\bf  Q }

\bibitem[104]{8850TP}
[104] Hao, L.-H. and Jiang, G.: 2011,
\newblock Phys. Rev. A 83, 012511,
\newblock {\bf  Q }

\bibitem[105]{8156TP}
[105] Hartman, H., Gurell, J., Lundin, P., Schef, P., Hibbert, A., Lundberg, H.,
  Mannervik, S., Norlin, L.-O., and Royen, P.: 2008,
\newblock Astron. Astrophys. 480, 575,
\newblock {\bf  LF, QF }

\bibitem[106]{8803TP}
[106] Hartman, H., Nilsson, H., Engstr\"om, L., Lundberg, H., Palmeri, P., Quinet,
  P., and Bi\'emont, E.: 2010,
\newblock Phys. Rev. A 82, 052512,
\newblock {\bf  L, Q }

\bibitem[107]{8845TP}
[107] Hartog, E. A.~D., Bilty, K.~A., and Lawler, J.~E.: 2011,
\newblock J. Phys. B 44, 055001,
\newblock {\bf  L }

\bibitem[108]{8578TP}
[108] Hartog, E. A.~D., Buettner, K.~P., and Lawler, J.~E.: 2009,
\newblock J. Phys. B 42, 085006,
\newblock {\bf  L }

\bibitem[109]{8846TP}
[109] Hartog, E. A.~D., Chisholm, J.~P., and Lawler, J.~E.: 2010,
\newblock J. Phys. B 43, 155004,
\newblock {\bf  L }

\bibitem[110]{8369TP}
[110] Hartog, E. A.~D. and Lawler, J.~E.: 2008,
\newblock J. Phys. B 41, 045701,
\newblock {\bf  L }

\bibitem[111]{14939EL}
[111] He, M., Therkildsen, K.~T., Jensen, B.~B., Brusch, A., Thomsen, J.~W., and
  Porsev, S.~G.: 2009,
\newblock Phys. Rev. A 80, 024501,
\newblock {\bf  IS, HFS }

\bibitem[112]{15655EL}
[112] Herrmann, Batteiger, M.~V., Kn\"unz~S., Saathoff~G., Udem~Th., and H\"ansch~T.~W., 2009,
\newblock Phys. Rev. Lett. 102, 013006 
\newblock {\bf  WL }

\bibitem[113]{8277TP}
[113] Hibbert, A. and Deb, N.~C.: 2008,
\newblock J. Phys.: Conf. Ser. 130, 012012,
\newblock {\bf  Q }

\bibitem[114]{8523TP}
[114] Hou, H.-J., Jiang, G., Hu, F., and Hao, L.-H.: 2009,
\newblock At. Data Nucl. Data Tables 95, 125,
\newblock {\bf  Q }

\bibitem[115]{8398TP}
[115] Hu, F., Jiang, G., Hong, W., and Hao, L.~H.: 2008,
\newblock Eur. Phys. J. D 49, 293,
\newblock {\bf  Q }

\bibitem[116]{8932TP}
[116] Hu, F., Jiang, G., Yang, J.~M., Zhang, J.~Y., and Zhao, X.~F.: 2011,
\newblock Acta Phys. Pol. A 120(3), 429,
\newblock {\bf  Q }

\bibitem[117]{8594TP}
[117] Hu, M.-H. and Wang, Z.-W.: 2009,
\newblock Chin. Phys. B 18, 2244,
\newblock {\bf  Q }

\bibitem[118]{8405TP}
[118] Hudson, C.~E.: 2009,
\newblock Astron. Astrophys. 493, 697,
\newblock {\bf  Q }

\bibitem[119]{14452EL}
[119] Ishikawa, Y., Encarnaci\'on, J. M.~L., and Tr\"abert, E.: 2009,
\newblock Phys. Scr. 79, 025301,
\newblock {\bf  Q }

\bibitem[120]{8283TP}
[120] Iskrenova-Tchoukova, E. and Safronova, M.~S.: 2008,
\newblock Phys. Rev. A 78, 012508,
\newblock {\bf  Q, QF }

\bibitem[121]{14098EL}
[121] Ivanov, T.~I., Salumbides, E.~J., Vieitez, M.~O., Cacciani, P.~C., de~Lange,
  C.~A., and Ubachs, W.: 2008,
\newblock Mon. Not. R. Astron. Soc. 389, L4,
\newblock {\bf  WL, CL }

\bibitem[122]{8930TP}
[122] Jensen, B.~B., Ming, H., Westergaard, P.~G., Gunnarsson, K., Madsen, M.~H.,
  Brusch, A., Hald, J., and Thomsen, J.~W.: 2011,
\newblock Phys. Rev. Lett. 107, 113001,
\newblock {\bf  LF }

\bibitem[123]{14471EL}
[123] Jin, W.-G., Nakai, H., Kawamura, M., and Minowa, T.: 2009a,
\newblock J. Phys. Soc. Jpn. 78, 015001,
\newblock {\bf  HFS, IS }

\bibitem[124]{14953EL}
[124] Jin, W.-G., Nemoto, Y., and Minowa, T.: 2009b,
\newblock J. Phys. Soc. Jpn. 78, 094301,
\newblock {\bf  IS, HFS }

\bibitem[125]{14408EL}
[125] Jin, W.-G., Nemoto, Y., Nakai, H., Kawamura, M., and Minowa, T.: 2008,
\newblock J. Phys. Soc. Jpn. 77, 124301,
\newblock {\bf  IS }

\bibitem[126]{14805EL}
[126] Johansson, S.: 2009,
\newblock Phys. Scr. T134, 014013,
\newblock {\bf  EL, CL }

\bibitem[127]{8717TP}
[127] J\"onsson, P., Li, J.-G., Gaigalas, G., and Dong, C.-Z.: 2010,
\newblock At. Data Nucl. Data Tables 96, 271,
\newblock {\bf  Q }

\bibitem[128]{8233TP}
[128] Kara{\c c}oban, B. and \"Ozdemir, L.: 2008a,
\newblock J. Quant. Spectrosc. Radiat. Transfer 109, 1968,
\newblock {\bf  Q }

\bibitem[129]{8152TP}
[129] Kara{\c c}oban, B. and \"Ozdemir, L.: 2008b,
\newblock Acta Phys. Pol. A 113(6), 1609,
\newblock {\bf  Q }

\bibitem[130]{8870TP}
[130] Kara{\c c}oban, B. and \"Ozdemir, L.: 2011,
\newblock J. Kor. Phys. Soc. 58, 417,
\newblock {\bf  Q }

\bibitem[131]{8776TP}
[131] Karmakar, S. and Das, M.~B.: 2010,
\newblock Eur. Phys. J. D 59, 361,
\newblock {\bf  L }

\bibitem[132]{8605TP}
[132] Karpu{\v s}kien\.e, R. and Bogdanovich, P.: 2009,
\newblock At. Data Nucl. Data Tables 95, 533,
\newblock {\bf  Q }

\bibitem[133]{8766TP}
[133] Kedzierski, D., Kusz, J., and Muzolf, J.: 2010,
\newblock Spectrochim. Acta, Part B 65, 248,
\newblock {\bf  TE }

\bibitem[134]{14158EL}
[134] Keenan, F.~P., Jess, D.~B., Aggarwal, K.~M., Thomas, R.~J., Brosius, J.~W., and
  Davila, J.~M.: 2008a,
\newblock Mon. Not. R. Astron. Soc. 389, 939,
\newblock {\bf  CL, Q, QF }

\bibitem[135]{15216EL}
[135] Keenan, F.~P., Milligan, R.~O., Jess, D.~B., Aggarwal, K.~M., Mathioudakis, M.,
  Thomas, R.~J., Brosius, J.~W., and Davila, J.~M.: 2010,
\newblock Mon. Not. R. Astron. Soc. 404, 1617,
\newblock {\bf  CL, WL }

\bibitem[136]{8199TP}
[136] Kelleher, D.~E. and Podobedova, L.~I.: 2008a,
\newblock J. Phys. Chem. Ref. Data 37, 709,
\newblock {\bf  CP }

\bibitem[137]{8363TP}
[137] Kelleher, D.~E. and Podobedova, L.~I.: 2008b,
\newblock J. Phys. Chem. Ref. Data 37, 1285,
\newblock {\bf  CP }

\bibitem[138]{8228TP}
[138] Kingston, A.~E. and Hibbert, A.: 2008,
\newblock J. Phys. B 41, 155001,
\newblock {\bf  Q }

\bibitem[139]{8649TP}
[139] Kingston, A.~E. and Hibbert, A.: 2009,
\newblock J. Phys. B 42, 185004,
\newblock {\bf  Q }

\bibitem[140]{8779TP}
[140] Kingston, A.~E. and Hibbert, A.: 2010,
\newblock J. Phys. B 43, 165003,
\newblock {\bf  Q }

\bibitem[141]{8625TP}
[141] Kisielius, R., Storey, P.~J., Ferland, G.~J., and Keenan, F.~P.: 2009,
\newblock Mon. Not. R. Astron. Soc. 397, 903,
\newblock {\bf  Q }

\bibitem[142]{8688TP}
[142] Kotochigova, S., Linnik, M., Kirby, K.~P., and Brickhouse, N.~S.: 2010,
\newblock Astrophys. J., Suppl. Ser. 186, 85,
\newblock {\bf  Q }

\bibitem[143]{12547EL}
[143] Kramida, A., Ryabtsev, A.~N., Ekberg, J.~O., Kink,~I., Mannervik,~S., Martinson,~I.: 2008,
\newblock Phys. Scr. 78, 025301
\newblock {\bf  CP }

\bibitem[144]{12548EL}
[144] Kramida, A., Ryabtsev, A.~N., Ekberg, J.~O., Kink,~I., Mannervik,~S., Martinson,~I.: 2008
\newblock Phys. Scr. 78, 025302 
\newblock {\bf  CP }

\bibitem[145]{16369EL}
[145] Kramida, A.: 2011a,
\newblock J. Res. Natl. Inst. Stand. Technol. 116(2), 599,
\newblock {\bf  EL, CL, WL }

\bibitem[146]{15291EL}
[146] Kramida, A.~E.: 2010a,
\newblock At. Data Nucl. Data Tables 96, 586,
\newblock {\bf  E, WL, CL, HFS, QF  }

\bibitem[147]{15340EL}
[147] Kramida, A.~E.: 2010b,
\newblock J. Phys. B 43, 205001,
\newblock {\bf  EL,WL, CL }

\bibitem[148]{15027EL}
[148] Krins, S., Oppel, S., Huet, N., von Zanthier, J., and Bastin, T.: 2009,
\newblock Phys. Rev. A 80, 062508,
\newblock {\bf  IS, HFS }

\bibitem[149]{15295EL}
[149] Kr\"oger, S., Er, A., \"Ozt\"urk, I.~K., Ba{\c s}ar, G., Jarmola, A., Ferber,
  R., Tamanis, M., and Za{\v c}s, L.: 2010,
\newblock Astron. Astrophys. 516, p. A70,
\newblock {\bf  HFS }

\bibitem[150]{8648TP}
[150] Ku{\l}aga-Egger, D. and Migda{\l}ek, J.: 2009,
\newblock J. Phys. B 42, 185002,
\newblock {\bf  Q }

\bibitem[151]{16262EL}
[151] Kumar, P. V.~K. and Suryanarayana, M.~V.: 2011,
\newblock J. Phys. B 44, 055003,
\newblock {\bf  IS, HFS, WL }

\bibitem[152]{8929TP}
[152] Landi, E.: 2011,
\newblock At. Data Nucl. Data Tables 97, 587,
\newblock {\bf  Q }

\bibitem[153]{8164TP}
[153] Landi, E. and Bhatia, A.~K.: 2008,
\newblock At. Data Nucl. Data Tables 94, 1,
\newblock {\bf  Q, QF }

\bibitem[154]{8524TP}
[154] Landi, E. and Bhatia, A.~K.: 2009a,
\newblock At. Data Nucl. Data Tables 95, 155,
\newblock {\bf  Q, QF }

\bibitem[155]{8630TP}
[155] Landi, E. and Bhatia, A.~K.: 2009b,
\newblock At. Data Nucl. Data Tables 95, 547,
\newblock {\bf  Q, QF }

\bibitem[156]{8689TP}
[156] Landi, E. and Bhatia, A.~K.: 2010,
\newblock At. Data Nucl. Data Tables 96, 52,
\newblock {\bf  Q, QF }

\bibitem[157]{14980EL}
[157] Landi, E. and Young, P.~R.: 2009a,
\newblock Astrophys. J. 706, 1,
\newblock {\bf  DB }

\bibitem[158]{15033EL}
[158] Landi, E. and Young, P.~R.: 2009b,
\newblock Astrophys. J. 707, 1191,
\newblock {\bf  CL, WL, EL,Q }

\bibitem[159]{15110EL}
[159] Landi, E. and Young, P.~R.: 2010a,
\newblock Astrophys. J. 713, 205,
\newblock {\bf  CL, WL, EL,Q }

\bibitem[160]{8877TP}
[160] Lawler, J.~E., Bilty, K.~A., and Hartog, E. A.~D.: 2011,
\newblock J. Phys. B 44, 095001,
\newblock {\bf  TE }

\bibitem[161]{8748TP}
[161] Lawler, J.~E., Chisholm, J., Nitz, D.~E., Wood, M.~P., Sobeck, J., and Hartog,
  E. A.~D.: 2010a,
\newblock J. Phys. B 43, 085701,
\newblock {\bf  TE }

\bibitem[162]{14909EL}
[162] Lawler, J.~E., Sneden, C., Cowan, J.~J., Ivans, I.~I., and Hartog, E. A.~D.:
  2009a,
\newblock Astrophys. J., Suppl. Ser. 182, 51,
\newblock {\bf  L, TE }

\bibitem[163]{8512TP}
[163] Lawler, J.~E., Sneden, C., Cowan, J.~J., Wyart, J.-F., Ivans, I.~I., Sobeck,
  J.~S., Stockett, M.~H., and Hartog, E. A.~D.: 2008b,
\newblock Astrophys. J., Suppl. Ser. 178, 71,
\newblock {\bf  TE }

\bibitem[164]{8807TP}
[164] Lawler, J.~E., Wyart, J.-F., and Hartog, E. A.~D.: 2010b,
\newblock J. Phys. B 43, 235001,
\newblock {\bf  TE }

\bibitem[165]{16803EL}
[165] Lee, S.~A. and Jr., W. M.~F.: 2010,
\newblock Phys. Rev. A 82, 042515,
\newblock {\bf  IS, HFS  }

\bibitem[166]{15609EL}
[166] Leefer, N., Cing\"oz, A., and Budker, D.: 2009,
\newblock Opt. Lett. 34, 2548,
\newblock {\bf  IS, HFS, EL }

\bibitem[167]{8397TP}
[167] Li, H.-L., Li, P., Cheng, Z., and Ma, H.-R.: 2008,
\newblock Commun. Theor. Phys. 49, 217,
\newblock {\bf  Q }

\bibitem[168]{8843TP}
[168] Liang, G.~Y. and Badnell, N.~R.: 2010,
\newblock Astron. Astrophys. 518, p. A64,
\newblock {\bf  Q }

\bibitem[169]{8805TP}
[169] Liang, G.~Y., Badnell, N.~R., L\'opez-Urrutia, J. R.~C., Baumann, T.~M., Zanna,
  G.~D., Storey, P.~J., Tawara, H., and Ullrich, J.: 2010,
\newblock Astrophys. J., Suppl. Ser. 190, 322,
\newblock {\bf  Q }

\bibitem[170]{14824EL}
[170] Liang, G.~Y., Baumann, T.~M., L\'opez-Urrutia, J. R.~C., Epp, S.~W., Tawara,
  H., Gonchar, A., Mokler, P.~H., Zhao, G., and Ullrich, J.: 2009a,
\newblock Astrophys. J. 696, 2275,
\newblock {\bf  CL }

\bibitem[171]{8593TP}
[171] Liang, G.~Y., Whiteford, A.~D., and Badnell, N.~R.: 2009b,
\newblock Astron. Astrophys. 499, 943,
\newblock {\bf  Q }

\bibitem[172]{8604TP}
[172] Liang, G.~Y., Whiteford, A.~D., and Badnell, N.~R.: 2009c,
\newblock Astron. Astrophys. 500, 1263,
\newblock {\bf  Q }

\bibitem[173]{15556EL}
[173] Liang, G.~Y. and Zhao, G.: 2010,
\newblock Mon. Not. R. Astron. Soc. 405, 1987,
\newblock {\bf  WL, CL }

\bibitem[174]{8247TP}
[174] Liang, L., Jiang, W.~X., Zhou, C., and Zhang, L.: 2008a,
\newblock Opt. Commun. 281, 2107,
\newblock {\bf  Q }

\bibitem[175]{8235TP}
[175] Liang, L. and Zhou, C.: 2008,
\newblock J. Quant. Spectrosc. Radiat. Transfer 109, 1995,
\newblock {\bf  Q }

\bibitem[176]{8358TP}
[176] Liang, L., Zhou, C., and Zhang, L.: 2008c,
\newblock Chin. Opt. Lett. 6, 804,
\newblock {\bf  Q }

\bibitem[177]{8788TP}
[177] L\'opez-Urrutia, J. R.~C. and Beiersdorfer, P.: 2010,
\newblock Astrophys. J. 721, 576,
\newblock {\bf  EF, LF }

\bibitem[178]{8916TP}
[178] Louzon, E., Feigel, A., Frank, Y., Raicher, E., Klapisch, M., Mandelbaum, P.,
  Levy, I., Hurvitz, G., Ehrlich, Y., Frankel, M., Maman, S., and Henis, Z.:
  2011,
\newblock High En. Dens. Phys. 7, 124,
\newblock {\bf  Q }

\bibitem[179]{8619TP}
[179] Luna, F. R.~T., Mania, A.~J., and Hernandes, J.~A.: 2009,
\newblock J. Appl. Spectrosc. 76, 447,
\newblock {\bf  Q }

\bibitem[180]{8750TP}
[180] Lundberg, H., Engstr\"om, L., Hartman, H., Nilsson, H., Palmeri, P., Quinet,
  P., and Bi\'emont, E.: 2010,
\newblock J. Phys. B 43, 085004,
\newblock {\bf  L, TE, Q }

\bibitem[181]{8309TP}
[181] Lundin, P., Gurell, J., Mannervik, S., Royen, P., Norlin, L.-O., Hartman, H.,
  and Hibbert, A.: 2008,
\newblock Phys. Scr. 78, 015301,
\newblock {\bf  LF, QF }

\bibitem[182]{8588TP}
[182] Malcheva, G., Mayo, R., Ortiz, M., Ruiz, P., Engstr\"om, L., Lundberg, H.,
  Nilsson, H., Quinet, P., Bi\'emont, E., and Blagoev, K.: 2009a,
\newblock Mon. Not. R. Astron. Soc. 395, 1523,
\newblock {\bf  L, Q }

\bibitem[183]{8874TP}
[183] Malcheva, G., Nilsson, H., Engstr\"om, L., Lundberg, H., Bi\'emont, E.,
  Palmeri, P., Quinet, P., and Blagoev, K.: 2011,
\newblock Mon. Not. R. Astron. Soc. 412, 1823,
\newblock {\bf  L, Q }

\bibitem[184]{8615TP}
[184] Malcheva, G., Yoca, S.~E., Mayo, R., Ortiz, M., Engstr\"om, L., Lundberg, H.,
  Nilsson, H., Bi\'emont, E., and Blagoev, K.: 2009b,
\newblock Mon. Not. R. Astron. Soc. 396, 2289,
\newblock {\bf  L, Q }

\bibitem[185]{8370TP}
[185] Mandal, S., Dixit, G., Sahoo, B.~K., Chaudhuri, R.~K., and Majumder, S.: 2008,
\newblock J. Phys. B 41, 055701,
\newblock {\bf  Q, QF }

\bibitem[186]{14969EL}
[186] Mania, A.~J., Luna, F. R.~T., Borges, F.~O., and Cavalcanti, G.~H.: 2009a,
\newblock J. Quant. Spectrosc. Radiat. Transfer 110, 2162,
\newblock {\bf  EL, CL, WL, Q }

\bibitem[187]{14359EL}
[187] Mania, A.~J., Luna, F. R.~T., and Hernandes, J.~A.: 2009c,
\newblock J. Quant. Spectrosc. Radiat. Transfer 110, 82,
\newblock {\bf  Q }

\bibitem[188]{8853TP}
[188] Mania, A.~J., Luna, F. R.~T., and Mania, E.: 2011,
\newblock J. Appl. Spectrosc. 77, 758,
\newblock {\bf  Q }

\bibitem[189]{8528TP}
[189] Mashonkina, L., Ryabchikova, T., Ryabtsev, A., and Kildiyarova, R.: 2009,
\newblock Astron. Astrophys. 495, 297,
\newblock {\bf  Q }

\bibitem[190]{15237EL}
[190] Mattolat, C., Gottwald, T., Raeder, S., Rothe, S., Schwellnus, F., Wendt, K.,
  Th\"orle-Pospiech, P., and Trautmann, N.: 2010,
\newblock Phys. Rev. A 81, 052513,
\newblock {\bf   EL }

\bibitem[191]{8561TP}
[191] Mel\'endez, J. and Barbuy, B.: 2009,
\newblock Astron. Astrophys. 497, 611,
\newblock {\bf  TE, Q }

\bibitem[192]{8875TP}
[192] Morton, D.~C. and Drake, G. W.~F.: 2011,
\newblock Phys. Rev. A 83, 042503,
\newblock {\bf  Q, QF }

\bibitem[193]{8338TP}
[193] Nahar, S.~N.: 2008,
\newblock J. Quant. Spectrosc. Radiat. Transfer 109, 2731,
\newblock {\bf  Q }

\bibitem[194]{8606TP}
[194] Nahar, S.~N.: 2009,
\newblock At. Data Nucl. Data Tables 95, 577,
\newblock {\bf  Q, QF }

\bibitem[195]{8691TP}
[195] Nahar, S.~N.: 2010a,
\newblock At. Data Nucl. Data Tables 96, 26,
\newblock {\bf  Q, QF }

\bibitem[196]{8785TP}
[196] Nahar, S.~N.: 2010b,
\newblock At. Data Nucl. Data Tables 96, 863,
\newblock {\bf  Q }

\bibitem[197]{8925TP}
[197] Nahar, S.~N.: 2011,
\newblock At. Data Nucl. Data Tables 97, 403,
\newblock {\bf  Q, QF }

\bibitem[198]{8521TP}
[198] Nahar, S.~N., Eissner, W., Sur, C., and Pradhan, A.~K.: 2009,
\newblock Phys. Scr. 79, 035401,
\newblock {\bf  Q, QF }

\bibitem[199]{8871TP}
[199] Nakhate, S.~G., Mukund, S., and Bhattacharyya, S.: 2010,
\newblock J. Quant. Spectrosc. Radiat. Transfer 111, 394,
\newblock {\bf  L }

\bibitem[200]{16364EL}
[200] Nave, G. and Sansonetti, C.~J.: 2011,
\newblock J. Opt. Soc. Am. B 28, 737,
\newblock {\bf  WL, EL }

\bibitem[201]{8361TP}
[201] Nilsson, H., Engstr\"om, L., Lundberg, H., Palmeri, P., Fivet, V., Quinet, P.,
  and Bi\'emont, E.: 2008,
\newblock Eur. Phys. J. D 49, 13,
\newblock {\bf  L, Q }

\bibitem[202]{8731TP}
[202] Nilsson, H., Hartman, H., Engstr\"om, L., Lundberg, H., Sneden, C., Fivet, V.,
  Palmeri, P., Quinet, P., and Bi\'emont, E.: 2010,
\newblock Astron. Astrophys. 511, p. A16,
\newblock {\bf  L, TE, Q, T }

\bibitem[203]{14265EL}
[203] Nilsson, H. and Ivarsson, S.: 2008a,
\newblock Astron. Astrophys. 492, 609,
\newblock {\bf  HFS, TE }

\bibitem[204]{15202EL}
[204] Nouri, Z., Rosner, S.~D., Li, R., Scholl, T.~J., and Holt, R.~A.: 2010,
\newblock Phys. Scr. 81, 065301,
\newblock {\bf  IS, HFS }

\bibitem[205]{8374TP}
[205] Oliver, P. and Hibbert, A.: 2008a,
\newblock J. Phys. B 41, 165003,
\newblock {\bf  Q }

\bibitem[206]{8279TP}
[206] Oliver, P. and Hibbert, A.: 2008b,
\newblock J. Phys.: Conf. Ser. 130, 012016,
\newblock {\bf  Q }

\bibitem[207]{8718TP}
[207] Oliver, P. and Hibbert, A.: 2010,
\newblock J. Phys. B 43, 074013,
\newblock {\bf  Q }

\bibitem[208]{8231TP}
[208] \"Ozdemir, L., \"Urer, G., and Kara{\c c}oban, B.: 2008,
\newblock J. Quant. Spectrosc. Radiat. Transfer 109, 1886,
\newblock {\bf  Q }

\bibitem[209]{8227TP}
[209] Palmeri, P., Quinet, P., Bi\'emont, E., Gurell, J., Lundin, P., Norlin, L.-O.,
  Royen, P., Blagoev, K., and Mannervik, S.: 2008,
\newblock J. Phys. B 41, 125703,
\newblock {\bf  LF, QF }

\bibitem[210]{8620TP}
[210] Palmeri, P., Quinet, P., Fivet, V., Bi\'emont, E., Cowley, C.~R., Engstr\"om,
  L., Lundberg, H., Hartman, H., and Nilsson, H.: 2009,
\newblock J. Phys. B 42, 165005,
\newblock {\bf  L, Q }

\bibitem[211]{9992TP}
[211] Palmeri, P., Quinet, P., Fivet, V., Bi\'emont, E., Nilsson, H., Engstr\"om,
  L., Lundberg, H.: 2008,
\newblock Phys. Scr. 78, 015304,
\newblock {\bf  L, Q }

\bibitem[212]{8841TP}
[212] Palmeri, P., Quinet, P., Mendoza, C., Bautista, M.~A., Garc\'ia, J., Witthoeft,
  M.~C., and Kallman, T.~R.: 2011,
\newblock Astron. Astrophys. 525, p. A59,
\newblock {\bf  Q }

\bibitem[213]{15273EL}
[213] Parthey, C.~G., Matveev, A., Alnis, J., Pohl, R., Udem, T., Jentschura, U.~D.,
  Kolachevsky, N., and H\"ansch, T.~W.: 2010,
\newblock Phys. Rev. Lett. 104, p. 233001,
\newblock {\bf  IS, EL, WL }

\bibitem[214]{8595TP}
[214] Podobedova, L.~I., Kelleher, D.~E., and Wiese, W.~L.: 2009,
\newblock J. Phys. Chem. Ref. Data 38, 171,
\newblock {\bf  CP }

\bibitem[215]{8920TP}
[215] Quinet, P., Bi\'emont, E., Palmeri, P., Engstr\"om, L., Hartman, H., Lundberg,
  H., and Nilsson, H.: 2011,
\newblock J. Electron Spectrosc. Relat. Phenom. 184, 174,
\newblock {\bf  Q }

\bibitem[216]{8406TP}
[216] Quinet, P., Fivet, V., Palmeri, P., Bi\'emont, E., Engstr\"om, L., Lundberg,
  H., and Nilsson, H.: 2009,
\newblock Astron. Astrophys. 493, 711,
\newblock {\bf  Q, L }

\bibitem[217]{8261TP}
[217] Quinet, P., Palmeri, P., Fivet, V., Bi\'emont, E., Nilsson, H., Engstr\"om, L.,
  and Lundberg, H.: 2008,
\newblock Phys. Rev. A 77, 022501,
\newblock {\bf  L, Q }

\bibitem[218]{8690TP}
[218] Ramsbottom, C.~A.: 2009,
\newblock At. Data Nucl. Data Tables 95, 910,
\newblock {\bf  Q }

\bibitem[219]{17707EL}
[219] Redman, S.~L., Lawler, J.~E., Nave, G., Ramsey, L.~W., and Mahadevan, S.: 2011,
\newblock Astrophys. J., Suppl. Ser. 195, 24,
\newblock {\bf  CL, WL }

\bibitem[220]{8673TP}
[220] Rehse, S.~J. and Ryder, C.~A.: 2009,
\newblock Spectrochim. Acta, Part B 64, 974,
\newblock {\bf  TE }

\bibitem[221]{8886TP}
[221] Safronova, M.~S. and Safronova, U.~I.: 2011a,
\newblock Phys. Rev. A 83, 052508,
\newblock {\bf  Q, QF }

\bibitem[222]{8329TP}
[222] Safronova, U.~I. and Mancini, R.: 2009,
\newblock At. Data Nucl. Data Tables 95, 54,
\newblock {\bf  Q }

\bibitem[223]{8349TP}
[223] Safronova, U.~I. and Safronova, M.~S.: 2008,
\newblock Phys. Rev. A 78, 052504,
\newblock {\bf  Q }

\bibitem[224]{8893TP}
[224] Safronova, U.~I. and Safronova, M.~S.: 2011b,
\newblock Can. J. Phys. 89, 465,
\newblock {\bf  Q, QF }

\bibitem[225]{8913TP}
[225] Sahoo, B.~K. and Das, B.~P.: 2011,
\newblock Phys. Rev. A 84, 012501,
\newblock {\bf  Q, QF }

\bibitem[226]{8600TP}
[226] Sahoo, B.~K., Das, B.~P., and Mukherjee, D.: 2009,
\newblock Phys. Rev. A 79, 052511,
\newblock {\bf  Q }

\bibitem[227]{8371TP}
[227] Sahoo, B.~K., Nataraj, H.~S., Das, B.~P., Chaudhuri, R.~K., and Mukherjee, D.:
  2008,
\newblock J. Phys. B 41, 055702,
\newblock {\bf  QF }

\bibitem[228]{15349EL}
[228] Saloman, E.~B.: 2010,
\newblock J. Phys. Chem. Ref. Data 39, 033101,
\newblock {\bf  CP }

\bibitem[229]{Salumbides}
[229] Salumbides, E.~J.~, Maslinskas, V., Dildar, U.~M., Wolf, A.~L., van Duijn, 
E.-J., Eikema, K.~S.~E., Ubachs, W.:  2011
\newblock Phys. Rev. A 83, 012502
\newblock {\bf  WL }

\bibitem[230]{17230EL}
[230] Sansonetti, C.~J., Simien, C.~E., Gillaspy, J.~D., Tan, J.~N., Brewer, S.~M.,
  Brown, R.~C., Wu, S.-J., and Porto, J.~V.: 2011,
\newblock Phys. Rev. Lett. 107, 023001,
\newblock {\bf  WL, HFS, IS, EL }

\bibitem[231]{15342EL}
[231] Sansonetti, C.~J. and Veza, D.: 2010,
\newblock J. Phys. B 43, 205003,
\newblock {\bf  WL, IS, HFS }

\bibitem[232]{8198TP}
[232] Sansonetti, J.~E.: 2008a,
\newblock J. Phys. Chem. Ref. Data 37, 7,
\newblock {\bf  CP }

\bibitem[233]{13237EL}
[233] Sansonetti, J.~E.: 2008b,
\newblock J. Phys. Chem. Ref. Data 37, 1659,
\newblock {\bf  CP }

\bibitem[234]{15024EL}
[234] Sansonetti, J.~E.: 2009a,
\newblock J. Phys. Chem. Ref. Data 38, 761,
\newblock {\bf  CP  }

\bibitem[235]{8682TP}
[235] Sansonetti, J.~E.: 2009b,
\newblock J. Phys. Chem. Ref. Data 38, 761,
\newblock {\bf  CP }

\bibitem[236]{15350EL}
[236] Sansonetti, J.~E. and Nave, G.: 2010a,
\newblock J. Phys. Chem. Ref. Data 39, 033103,
\newblock {\bf  CP }

\bibitem[237]{8917TP}
[237] Santos, J.~P., Costa, A.~M., Madruga, C., Parente, F., and Indelicato, P.:
  2011,
\newblock Eur. Phys. J. D 63, 89,
\newblock {\bf  Q }

\bibitem[238]{8742TP}
[238] Shah, M.~L., Pulhani, A.~K., Gupta, G.~P., and Suri, B.~M.: 2010,
\newblock J. Opt. Soc. Am. B 27, 423,
\newblock {\bf  L, TE }

\bibitem[239]{8753TP}
[239] Shen, X.-Z., Yuan, P., and Liu, J.: 2010,
\newblock Chin. Phys. B 19, 053101,
\newblock {\bf  Q, QF }

\bibitem[240]{14040EL}
[240] Shestov, S.~V., Bozhenkov, S.~A., Zhitnik, I.~A., Kuzin, S.~V., Urnov, A.~M.,
  Beigman, I.~L., Goryaev, F.~F., and Tolstikhina, I.~Y.: 2008,
\newblock Astron. Lett. 34, 33,
\newblock {\bf  CL, WL }

\bibitem[241]{8762TP}
[241] Singh, J., Jha, A. K.~S., and Mohan, M.: 2010a,
\newblock J. Phys. B 43, 115005,
\newblock {\bf  Q }

\bibitem[242]{8789TP}
[242] Singh, J., Jha, A. K.~S., Verma, N., and Mohan, M.: 2010b,
\newblock At. Data Nucl. Data Tables 96, 759,
\newblock {\bf  Q }

\bibitem[243]{14097EL}
[243] Smillie, D.~G., Pickering, J.~C., and Smith, P.~L.: 2008,
\newblock Mon. Not. R. Astron. Soc. 390, 733,
\newblock {\bf  WL, CL }

\bibitem[244]{15487EL}
[244] Steenstrup, M.~P., Brusch, A., Jensen, B.~B., Hald, J., and Thomsen, J.~W.:
  2010,
\newblock Phys. Rev. A 82, 054501,
\newblock {\bf  IS }

\bibitem[245]{8715TP}
[245] Storey, P.~J. and Zeippen, C.~J.: 2010,
\newblock Astron. Astrophys. 511, p. A78,
\newblock {\bf  Q }

\bibitem[246]{15043EL}
[246] Tauheed, A., Joshi, Y.~N., and Steinitz, M.: 2009,
\newblock Can. J. Phys. 87, 1255,
\newblock {\bf  WL, CL, EL }

\bibitem[247]{8159TP}
[247] Tayal, S.~S.: 2008a,
\newblock Astron. Astrophys. 486, 629,
\newblock {\bf  Q }

\bibitem[248]{8356TP}
[248] Tayal, S.~S.: 2008b,
\newblock Astrophys. J., Suppl. Ser. 178, 359,
\newblock {\bf  Q }

\bibitem[249]{8518TP}
[249] Tayal, S.~S.: 2009a,
\newblock Phys. Scr. 79, 015303,
\newblock {\bf  Q }

\bibitem[250]{8660TP}
[250] Tayal, S.~S.: 2009b,
\newblock Phys. Rev. A 80, 032512,
\newblock {\bf  QF }

\bibitem[251]{8851TP}
[251] Tayal, S.~S.: 2011a,
\newblock Phys. Rev. A 83, 012515,
\newblock {\bf  Q, QF }

\bibitem[252]{8921TP}
[252] Tayal, S.~S.: 2011b,
\newblock At. Data Nucl. Data Tables 97, 481,
\newblock {\bf  Q }

\bibitem[253]{8758TP}
[253] Tayal, S.~S. and Zatsarinny, O.: 2010a,
\newblock Astrophys. J., Suppl. Ser. 188, 32,
\newblock {\bf  Q, QF }

\bibitem[254]{8714TP}
[254] Tayal, S.~S. and Zatsarinny, O.: 2010b,
\newblock Astron. Astrophys. 510, p. A79,
\newblock {\bf  Q }

\bibitem[255]{8656TP}
[255] Tayal, V., Gupta, G.~P., and Tripathi, A.~N.: 2009,
\newblock Indian J. Phys. 83, 1271,
\newblock {\bf  Q }

\bibitem[256]{8577TP}
[256] Therkildsen, K.~T., Jensen, B.~B., Ryder, C.~P., Malossi, N., and Thomsen,
  J.~W.: 2009,
\newblock Phys. Rev. A 79, 034501,
\newblock {\bf  TE }

\bibitem[257]{15916EL}
[257] Thorne, A.~P., Pickering, J.~C., and Semeniuk, J.: 2011,
\newblock Astrophys. J., Suppl. Ser. 192, 11,
\newblock {\bf  CL, WL, EL }

\bibitem[258]{8281TP}
[258] Tr\"abert, E., Hoffman, J., Reinhardt, S., Wolf, A., and Zanna, G.~D.: 2008,
\newblock J. Phys.: Conf. Ser. 130, 012018,
\newblock {\bf  LF }

\bibitem[259]{8336TP}
[259] Tr\"abert, E., Hoffmann, J., Krantz, C., Wolf, A., Ishikawa, Y., and Santana,
  J.~A.: 2009,
\newblock J. Phys. B 42, 025002,
\newblock {\bf  LF }

\bibitem[260]{17183EL}
[260] Tr\"abert, E., Ishikawa, Y., Santana, J.~A., and Zanna, G.~D.: 2011,
\newblock Can. J. Phys. 89, 403,
\newblock {\bf  EL, CL, TE }

\bibitem[261]{14931EL}
[261] Uddin, Z. and Windholz, L.: 2009,
\newblock Chin. J. Phys. 47(4), 454,
\newblock {\bf  HFS, CL, WL }

\bibitem[262]{14866EL}
[262] Wallace, L. and Hinkle, K.: 2009,
\newblock Astrophys. J. 700, 720,
\newblock {\bf  CL, WL, EL }

\bibitem[263]{8176TP}
[263] Wang, Z.-W., Li, X.-R., Hu, M.-H., Liu, Y., and Wang, Y.-N.: 2008a,
\newblock Chin. Phys. Lett. 25, 2004,
\newblock {\bf  Q }

\bibitem[264]{8359TP}
[264] Wang, Z.-W., Liu, Y., Hu, M.-H., Li, X.-R., and Wang, Y.-N.: 2008b,
\newblock Chin. Phys. B 17, 2909,
\newblock {\bf  Q }

\bibitem[265]{8354TP}
[265] Wang, Z.-W., Wang, Y.-N., Hu, M.-H., Li, X.-R., and Liu, Y.: 2008c,
\newblock Sci. China, Ser. G 51, 1633,
\newblock {\bf  Q }

\bibitem[266]{14846EL}
[266] Wasowicz, T.~J.: 2009,
\newblock Eur. Phys. J. D 53, 263,
\newblock {\bf  IS }

\bibitem[267]{15501EL}
[267] Wasowicz, T.~J., Werbowy, S., Kwela, J., and Drozdowski, R.: 2010,
\newblock J. Opt. Soc. Am. B 27, 2628,
\newblock {\bf  IS }

\bibitem[268]{8823TP}
[268] Wei, H.~G., Shi, J.~R., Zhao, G., and Liang, Z.~T.: 2010,
\newblock Astron. Astrophys. 522, p. A103,
\newblock {\bf  Q }

\bibitem[269]{8610TP}
[269] Wiese, W.~L. and Fuhr, J.~R.: 2009,
\newblock J. Phys. Chem. Ref. Data 38, 565,
\newblock {\bf  CP }

\bibitem[270]{8316TP}
[270] Witthoeft, M.~C. and Badnell, N.~R.: 2008,
\newblock Astron. Astrophys. 481, 543,
\newblock {\bf  QF }

\bibitem[271]{14130EL}
[271] Wolf, A.~L., van~den Berg, S.~A., Gohle, C., Salumbides, E.~J., Ubachs, W., and
  Eikema, K. S.~E.: 2008,
\newblock Phys. Rev. A 78, 032511,
\newblock {\bf  WL, EL }

\bibitem[272]{14601EL}
[272] Wyart, J.-F. and Lawler, J.~E.: 2009a,
\newblock Phys. Scr. 79, 045301,
\newblock {\bf  EL, CL, W, Q }

\bibitem[273]{8157TP}
[273] Wyart, J.-F., Tchang-Brillet, W.-U.~L., Churilov, S.~S., and Ryabtsev, A.~N.:
  2008,
\newblock Astron. Astrophys. 483, 339,
\newblock {\bf  Q }

\bibitem[274]{15917EL}
[274] Xie, J., Dai, C.-J., and Li, M.: 2011,
\newblock J. Phys. B 44, 015002,
\newblock {\bf  EL }

\bibitem[275]{8647TP}
[275] Xu, J.-X., Feng, Y.-Y., Sun, G.-J., and Dai, Z.-W.: 2009,
\newblock Chin. Phys. B 18, 3828,
\newblock {\bf  L }

\bibitem[276]{8612TP}
[276] y.~Zhang, T. and w.~Zheng, N.: 2009,
\newblock Chin. J. Chem. Phys. 22, 246,
\newblock {\bf  Q }

\bibitem[277]{8331TP}
[277] Yang, J.-H., Li, P., Zhang, J.-P., and Li, H.-L.: 2008,
\newblock Commun. Theor. Phys. 50, 468,
\newblock {\bf  Q }

\bibitem[278]{16371EL}
[278] Yang, Z.~H., Du, S.~B., Chang, H.~W., Zhang, Y.~P., Zhang, B.~L., Xu, Q.~M.,
  Yu, D.~Y., and Cai, X.~H.: 2010,
\newblock J. Quant. Spectrosc. Radiat. Transfer 111, 2007,
\newblock {\bf  WL, CL }

\bibitem[279]{14522EL}
[279] Yang, Z.~H., Du, S.~B., Zeng, X.~T., Chang, H.~W., Zhang, B.~L., Wang, W., Yu,
  D.~Y., and Cai, X.~H.: 2009,
\newblock Astron. J. 137, 4020,
\newblock {\bf  CL, WL }

\bibitem[280]{8573TP}
[280] Yildiz, M., {\c C}elik, G., and Kili{\c c}, H.~{\c S}.: 2009,
\newblock Acta Phys. Pol. A 115, 641,
\newblock {\bf  Q }

\bibitem[281]{8310TP}
[281] Yoca, S.~E., Bi\'emont, E., Delahaye, F., Quinet, P., and Zeippen, C.~J.: 2008,
\newblock Phys. Scr. 78, 025303,
\newblock {\bf  Q }

\bibitem[282]{14442EL}
[282] Young, P.~R.: 2009,
\newblock Astrophys. J. 691, L77,
\newblock {\bf  CL, WL, EL }

\bibitem[283]{17711EL}
[283] Young, P.~R., Feldman, U., and Lobel, A.: 2011,
\newblock Astrophys. J., Suppl. Ser. 196, 23,
\newblock {\bf  CL, WL, EL }

\bibitem[284]{15032EL}
[284] Young, P.~R. and Landi, E.: 2009,
\newblock Astrophys. J. 707, 173,
\newblock {\bf  CL, WL, EL }

\bibitem[285]{17710EL}
[285] Y\"uce, K., Castelli, F., and Hubrig, S.: 2011a,
\newblock Astron. Astrophys. 528, p. A37,
\newblock {\bf  CL, WL, EL, TE }

\bibitem[286]{15001EL}
[286] Zanna, G.~D.: 2009a,
\newblock Astron. Astrophys. 508, 501,
\newblock {\bf  CL, WL, EL, Q }

\bibitem[287]{15002EL}
[287] Zanna, G.~D.: 2009c,
\newblock Astron. Astrophys. 508, 513,
\newblock {\bf  CL, WL, EL, Q }

\bibitem[288]{15195EL}
[288] Zanna, G.~D.: 2010a,
\newblock Astron. Astrophys. 514, p. A41,
\newblock {\bf  EL, CL, WL, Q, QF }

\bibitem[289]{15022EL}
[289] Zanna, G.~D. and Ishikawa, Y.: 2009,
\newblock Astron. Astrophys. 508, 1517,
\newblock {\bf  EL, CL, WL }

\bibitem[290]{8319TP}
[290] Zanna, G.~D., Rozum, I., and Badnell, N.~R.: 2008,
\newblock Astron. Astrophys. 487, 1203,
\newblock {\bf  Q, QF }

\bibitem[291]{15194EL}
[291] Zanna, G.~D., Storey, P.~J., and Mason, H.~E.: 2010a,
\newblock Astron. Astrophys. 514, p. A40,
\newblock {\bf  EL, CL, WL, Q }

\bibitem[292]{8642TP}
[292] Zhang, T.-Y. and Zheng, N.-W.: 2009,
\newblock Acta Phys. Pol. A 116(2), 141,
\newblock {\bf  Q }

\bibitem[293]{8333TP}
[293] Zhang, T.-Y., Zheng, N.-W., and Ma, D.-X.: 2009,
\newblock Int. J. Quantum Chem. 109, 145,
\newblock {\bf  Q }

\bibitem[294]{17164EL}
[294] Zhang, W., Du, S., Feng, Y.-Y., Jiang, L.-Y., Jiang, Z.-K., and Dai, Z.-W.:
  2011a,
\newblock Mon. Not. R. Astron. Soc. 413, 1803,
\newblock {\bf  L }

\bibitem[295]{8795TP}
[295] Zhang, W., Feng, Y.-Y., and Dai, Z.-W.: 2010a,
\newblock J. Opt. Soc. Am. B 27, 2255,
\newblock {\bf  L }

\bibitem[296]{8808TP}
[296] Zhang, W., Feng, Y.-Y., Sun, G.-J., and Dai, Z.-W.: 2010b,
\newblock J. Phys. B 43, 235005,
\newblock {\bf  L }

\bibitem[297]{8790TP}
[297] Zhang, W., Feng, Y.-Y., Xu, J.-X., Palmeri, P., Quinet, P., Bi\'emont, E., and
  Dai, Z.-W.: 2010c,
\newblock J. Phys. B 43, 205005,
\newblock {\bf  L }

\bibitem[298]{15462EL}
[298] Zhang, W., Palmeri, P., Quinet, P., Bi\'emont, E., Du, S., and Dai, Z.-W.:
  2010d,
\newblock Phys. Rev. A 82, 042507,
\newblock {\bf  EL, CL, L }

\bibitem[299]{12817EL}
[299] Zhang, Y., Xu, J.-X., Zhang, W., You, S., Ma, Z.-G., Han, L.-L., Li, P.-F.,
  Sun, G.-J., Jiang, Z.-K., Yoca, S.~E., Quinet, P., Bi\'emont, E., and Dai,
  Z.-W.: 2008a,
\newblock Phys. Rev. A 78, 022505,
\newblock {\bf  L  }

\end{thebibliography}

\end{document}